\documentclass[]{article}
\makeatletter\if@twocolumn\PassOptionsToPackage{switch}{lineno}\else\fi\makeatother

\usepackage{amsfonts,amssymb,amsbsy,latexsym,amsmath,tabulary,graphicx,times,caption,fancyhdr}
\usepackage[utf8]{inputenc}

\usepackage{url,multirow,morefloats,floatflt,cancel,tfrupee}
\makeatletter

\usepackage{setspace}
\usepackage{lineno}

\AtBeginDocument{\@ifpackageloaded{textcomp}{}{\usepackage{textcomp}}}
\makeatother
\usepackage{colortbl}
\usepackage{pifont}
\usepackage[table]{xcolor}
\usepackage{tablestyles}
\usepackage{longtable}
\usepackage[nointegrals]{wasysym}
\usepackage{hyperref}
\usepackage{tabularx,ragged2e,rotating}
 \usepackage{booktabs}
 \usepackage{pdflscape} 
\newcolumntype{C}{>{\hspace{0pt}%
    \Centering\arraybackslash}X}
\usepackage[skip=0.333\baselineskip]{caption}
\usepackage{subfig}
\usepackage{doi}
\makeatletter

\def\mcWidth#1{\csname TY@F#1\endcsname+\tabcolsep}

\def\cAlignHack{\rightskip\@flushglue\leftskip\@flushglue\parindent\z@\parfillskip\z@skip}
\def\rAlignHack{\rightskip\z@skip\leftskip\@flushglue \parindent\z@\parfillskip\z@skip}

\@ifundefined{etal}{}{}

\usepackage{ifxetex}
\ifxetex\else\if@twocolumn\@ifpackageloaded{stfloats}{}{\usepackage{dblfloatfix}}\fi\fi

\AtBeginDocument{
\expandafter\ifx\csname eqalign\endcsname\relax
\def\eqalign#1{\null\vcenter{\def\\{\cr}\openup\jot\m@th
  \ialign{\strut$\displaystyle{##}$\hfil&$\displaystyle{{}##}$\hfil
      \crcr#1\crcr}}\,}
\fi
}

\AtBeginDocument{%
  \@ifpackageloaded{endfloat}%
   {\renewcommand\efloat@iwrite[1]{\immediate\expandafter\protected@write\csname efloat@post#1\endcsname{}}}{\newif\ifefloat@tables}%
}%

\def\BreakURLText#1{\@tfor\brk@tempa:=#1\do{\brk@tempa\hskip0pt}}
\let\lt=<
\let\gt=>
\def\processVert{\ifmmode|\else\textbar\fi}

\@ifundefined{subparagraph}{
\def\subparagraph{\@startsection{paragraph}{5}{2\parindent}{0ex plus 0.1ex minus 0.1ex}%
{0ex}{\normalfont\small\itshape}}%
}{}

\newcommand\role[1]{\unskip}
\newcommand\aucollab[1]{\unskip}
  
\@ifundefined{tsGraphicsScaleX}{\gdef\tsGraphicsScaleX{1}}{}
\@ifundefined{tsGraphicsScaleY}{\gdef\tsGraphicsScaleY{.9}}{}
\def\checkGraphicsWidth{\ifdim\Gin@nat@width>\linewidth
	\tsGraphicsScaleX\linewidth\else\Gin@nat@width\fi}

\def\checkGraphicsHeight{\ifdim\Gin@nat@height>.9\textheight
	\tsGraphicsScaleY\textheight\else\Gin@nat@height\fi}

\def\fixFloatSize#1{}
\let\ts@includegraphics\includegraphics

\def\inlinegraphic[#1]#2{{\edef\@tempa{#1}\edef\baseline@shift{\ifx\@tempa\@empty0\else#1\fi}\edef\tempZ{\the\numexpr(\numexpr(\baseline@shift*\f@size/100))}\protect\raisebox{\tempZ pt}{\ts@includegraphics{#2}}}}

\AtBeginDocument{\def\includegraphics{\@ifnextchar[{\ts@includegraphics}{\ts@includegraphics[width=\checkGraphicsWidth,height=\checkGraphicsHeight,keepaspectratio]}}}

\DeclareMathAlphabet{\mathpzc}{OT1}{pzc}{m}{it}

\def\URL#1#2{\@ifundefined{href}{#2}{\href{#1}{#2}}}

\def\UrlOrds{\do\*\do\-\do\~\do\'\do\"\do\-}%
\g@addto@macro{\UrlBreaks}{\UrlOrds}

\edef\fntEncoding{\f@encoding}

\makeatother

\newif\ifmultipleabstract\multipleabstractfalse%
\makeatletter

\def\wileyIndent{1pt}
\usepackage[paperheight=10in,paperwidth=6.5in,margin=2cm,headsep=.5cm,top=2.5cm,headheight=1cm]{geometry}

\renewenvironment{abstract}
{\vspace*{-1pc}\trivlist\item[]\leftskip\wileyIndent\hrulefill\par\vskip4pt\noindent\textbf{\abstractname}\mbox{\null}\\}{\par\noindent\hrulefill\endtrivlist}

\usepackage[]{footmisc}

\def\author#1{\gdef\@author{\hskip-\dimexpr(\tabcolsep)\hskip\wileyIndent\parbox{\dimexpr\textwidth-\wileyIndent}{\centering\bfseries#1}}}

\def\title#1{\linespread{1}\gdef\@title{\centering\bfseries\ifx\@articleType\@empty\else\@articleType\\\fi#1}}

\let\@articleType\@empty \def\articletype#1{\gdef\@articleType{{\normalfont\itshape#1}}}

\AtBeginDocument{\fancypagestyle{headings}{\fancyhf{}\fancyhead[C]{\RunningHead}\fancyhead[R]{\thepage}}\pagestyle{headings}}

 \def\audegree#1{}

\date{}

\makeatother

\usepackage[T1]{fontenc}
\makeatother
\usepackage[sort,compress]{natbib}
\usepackage{notoccite}

\def\thanksspace{{\phantom{\textsuperscript{\thefootnote}}}}

\begin{document}


\title{Modelling heterogeneity in the classification process in multi-species distribution models can improve predictive performance.}
\author{Kwaku Peprah~Adjei\textsuperscript{1,2}\thanks{Corresponding author.}\thanksspace \thanks{E-mail:                     
                   kwaku.p.adjei@ntnu.no}{\thanksspace}\space Robert B.~O'Hara\textsuperscript{1,2}\thanks{E-mail:        bob.ohara@ntnu.no}{\thanksspace} \space Wouter Koch\textsuperscript{2,4}\thanks{E-mail:                 wouter.koch@artsdatabanken.no}{\thanksspace} and Anders Finstad \textsuperscript{2,3}\thanks{E-mail:anders.finstad@ntnu.no}{\thanksspace}~\\[-3pt]\normalsize\normalfont  \itshape ~\\
\textsuperscript{1}{Department of Mathematical Sciences\unskip, Norwegian University of Science and Technology\unskip, Trondheim\unskip, Norway}~\\
\textsuperscript{2}{Center for Biodiversity Dynamics\unskip, Norwegian University of Science and Technology\unskip, Norway}~\\
\textsuperscript{3}{Department of Natural History \unskip, Norwegian University of Science and Technology\unskip, Norway}~\\
\textsuperscript{4}{Norwegian Biodiversity Information Center \unskip, Trondheim \unskip, Norway}}

\def\RunningHead{Accounting for heterogeneity in misclassification process }\def\RunningAuthor{Adjei}

\maketitle 

\doublespacing


\begin{abstract}
   \begin{enumerate}
\item Species distribution models and maps from large-scale biodiversity data are necessary for conservation management. One current issue is that biodiversity data are prone to taxonomic misclassifications. Methods to account for these misclassifications in multispecies distribution models have assumed that the classification probabilities are constant throughout the study. In reality, classification probabilities are likely to vary with several covariates. Failure to account for such heterogeneity can lead to bias in parameter estimates. 

\item Here we present a general multispecies distribution model that accounts for heterogeneity in the classification process. The proposed model assumes a multinomial generalised linear model for the classification confusion matrix. We compare the performance of the heterogeneous classification model to that of the homogeneous classification model by assessing how well they estimate the parameters in the model and their predictive performance on hold-out samples. We applied the model to gull data from Norway, Denmark and Finland, obtained from Global Biodiversity Information Facility.

\item Our simulation study showed that accounting for heterogeneity in the classification process increased precision by 30 \% and reduced accuracy and recall by 6\%. Applying the model framework to the gull dataset did not improve the predictive performance between the homogeneous and heterogeneous models due to the smaller misclassified sample sizes. However, when machine learning predictive scores are used as weights to inform the species distribution models about the classification process, the precision increases by 70\%. 

\item We recommend multiple multinomial regression to be used to model the variation in the classification process when the data contains relatively larger misclassified samples. Machine prediction scores should be used when the data contains relatively smaller misclassified samples. 
    \end{enumerate}
    
\def\keywordstitle{Keywords}

\end{abstract}

\section{Introduction}

Species distribution models are essential ecology and conservation management tools that predict how natural and human factors affect biodiversity \citep{elith2009species,vermeiren2020integrating}. With an  increasing amount of biodiversity data from multi-species surveys available to scientists, multi-species distribution models (hereafter mSDMs) have become widely used in analysing these data. These mSDMs  model data at the community level, identify the important variables that drive species co-occurrences and predict the distribution of species in a community \citep{ovaskainen2011making, pollock2014understanding, hui2015multi}. 

However, the biodiversity data obtained from these surveys can be subject to observation errors. These observation errors include false positives (recording absent species as present and misidentifying one species for another) and false negatives (failing to record present species). These false positives are usually a species that has been misclassified. It would be a great advantage if the observation was correctly classified, rather than being discarded once it has been identified as a false positive. The false positives may arise from imperfect classifiers  \citep{wright2020modelling, spiers2022estimating}, observer error, and many other sources. In this study, we collectively describe all the false positives and misidentifications as misclassification. We use the term true states in this study to describe the correct or actual observation identity we are interested in modelling. False negatives are mostly accounted for in occupancy models by modelling the imperfect detection in the observation model. Failure to account for or correct these errors leads to biases in inferences about state variables such as occupancy probabilities, covariate effects and relative activity \citep{ ferguson2015occupancy, wright2020modelling, clare2021generalized}, leading to an impairment in decision making \citep{hoekman2021multi}. 

The methods to deal with misclassification from biodiversity data can be grouped into data review methods and model-based methods \citep{clare2021generalized}. Data review methods require complete and proper data collection and processing methods. This process can be very demanding as it is challenging to control for misclassification. This makes the model-based methods more popular when working with large-scale datasets from large-scale biodiversity data vendors like the Global Biodiversity Information Facility \citep[GBIF hereafter;][]{gbif}. Model-based methods estimate classification probabilities jointly with the true state variables of interest. Model-based methods attempting to account for misclassification in multispecies occupancy models currently include modelling misclassification with detection heterogeneity \citep{ferguson2015occupancy, louvrier2018accounting,clement2022estimating}, integrating multiple observers records with other methods such as distance sampling and  N-Mixture models \citep{hoekman2021multi}, supervised methods with extra information from observation confirmation or verification \citep{ferguson2015occupancy, guillera2017dealing}, site confirmation \citep{clare2021generalized} and other calibrated methods. These methods need extra data from the verification process, which helps in estimating the misclassification probabilities in a semi-supervised setting \citep{spiers2022estimating} and makes the parameters in the model identifiable \citep{guillera2017dealing}. The above-mentioned studies have either used verified data collected on the site-level \citep[where the occupancy state of a species is known at a site and not at the individual sample level; ][]{chambert2018new}, on aggregated individual sample level using a multinomial model with site-covariates \citep{wright2020modelling}, or on individual sample-level validation data which helps in modelling non-species identities (morphospecies) to species identities \citep{spiers2022estimating}. It is also worth stating that some studies have explored misclassification in abundance \citep{conn2013accommodating} and capture-recapture \citep{augustine2020spatial} models.\\

Furthermore, these previous studies assumed that the misclassification probabilities are homogeneous (constant) across the study. In reality, the classification probabilities may vary with environmental covariates \citep[such as field conditions;][]{conn2013accommodating} or observer experience \citep[especially when ascertaining how well each observer classifies a report in citizen science projects will be informative;][]{ arazy2021framework, johnston2022outstanding}, distance from a transect when using transect data \citep{conn2013accommodating}, picture quality, etc. An attempt at modelling the heterogeneity in the classification process would be to model this effect as part of the true state model (such as abundance or occupancy). However, this approach may not solve the problem of heterogeneity in the classification process since the estimates of the true state observation process parameters only serve as informed priors to the classification process \citep{spiers2022estimating}. 

A more correct approach to model this heterogeneity is adding the covariate effect to the classification process. Some studies on dynamic false positive single-species occupancy models have modelled temporal changes in false positives using year as a covariate \citep{sutherland2013accounting, miller2013determining, kery2020applied}, showing the possibility to model misclassification trends over time. Our study attempts to model variation in classification probabilities in mSDMs by modelling the probability of classifying a true state identity with a multinomial generalised linear model. To our knowledge, no previous work has been done on this. Failure to account for the heterogeneity in the misclassification probabilities will lead to biased estimates in the parameters of the true state observation process (such as species abundance, richness, and occupancy probabilities) and reduce the model's predictive performance. 

Moreover, recent efforts to correctly classify observations from biodiversity surveys have relied on machine learning (hereafter ML) algorithms  \citep{suzuki2022deep, saoud2020miss, lotfian2021partnership, willi2019identifying,keshavan2019combining, borowiec2022deep, koch2022maximizing}. These ML algorithms use sounds and/or images of observations to predict the true identity of the individual observations, and they can be trained to mimic expert verification of observations \citep{keshavan2019combining, ponti2022human, langenkamper2019impact}. These ML algorithms use a prediction score (a value that shows the weight of predicting the observations as something else) to predict the possible list of the true identities of the individual reported observations. These prediction scores and a list of possible true identities provide information about the classification process of each observation. They can be used to model heterogeneity in the classification process. This study is the first to model the heterogeneity in the classification process by using the prediction scores to weigh the distribution of the reported observations and predict the distribution of the actual observation identities. 

Here, we present a joint model that simultaneously models the true state variables of interest (such as occupancy and abundance) and the heterogeneity in the classification process. Our model set-up extends the work done by \cite{wright2020modelling} and \cite{spiers2022estimating} by a) allowing the classification probabilities to vary with a covariate, b) allowing additional morphostates designations in imperfect classifications (for example, having "unidentified woodpecker" as part of the classified observations), c) using ML prediction scores as weights to account for heterogeneity in the classification process and d) performing variable selection on the classification process covariate to check for potential mSDM overfit. We compared the classification performance of our model with models that assume a homogeneous classification probability done by \cite{wright2020modelling} and \citep{spiers2022estimating} through simulation studies. We parameterise our model  with citizen science data on gulls in Norway, Finland and Denmark from iNaturalist \citep{matheson2014inaturalist} downloaded from the GBIF \citep{gbif}. 

\section{Methodology}
\label{section:methodology}

\subsection{Model framework}
The proposed framework assumes a model for where the true state's identity is "present" (we use "state(s)" in this work to refer to taxon identity as well as any other identification category or morpho-states, such as an unidentified group). This true state can be on any taxonomic level. However, the true states are often unknown and treated as latent. An individual is sampled and classified as another state, known as the "reported" or "classified" state, with a probability. These reported states can also be on any taxonomic level or any unidentified group \citep[such as in studies that include morphospecies;][]{spiers2022estimating}. We further assume that the individuals are verifiable (and that we have information on the verification process) and that the verified state approximates the true state identity. 

\subsubsection{True state observation process model}

We specify a relative abundance model (specifically a multinomial logit model) for the true state observation process. Our objective is to show how to model heterogeneity in the classification process and not to make inferences about the true state's abundance, so we chose a model that was easier to fit and understand to explain the distribution of the true state.  

Let $\lambda_{is}$ be the average number of $s = 1,2,\ldots, S$ true individuals  at location $i =1,2,\ldots, R$, which describes the abundance of the true states over space.  This means that intensity can be modelled as an inhomogeneous point process (PPM) which assumes that the data are dependent on the environment covariate, or as a log-Cox Gaussian Point process (LGCP) where we assume a spatial dependency in the observed data \citep{renner2015point}. Here, the mean intensity is modelled using the inhomogeneous  point process and defined as:

\begin{equation} \label{intensity}
 \ln (\lambda_{is})= \beta_{0s} + \sum_{j = 1}^{n_e} x_{ij}\beta_{js}  ,
\end{equation}
where $\mathbf{\beta_{0s}}$ is the intercept of state $s$, $\beta_{js}$ is the effect of covariate $j$ on the intensity of true state $s$, $x_{ij}$ is the $j^{th}$ covariate that affects the observation of the state at location $i$ and $n_e$ is the number of covariates in the observation process model. Note that we assume there are no species interactions in our relative abundance model, and this could have been added as a random effect in the true state intensity definition (equation \eqref{intensity}).

Let $p_{is}$ be the probability of true state $s$ being at location $i$. We estimate this probability from the mean intensities as follows:

\begin{equation}\label{proportion}
    p_{is} = \frac{\lambda_{is}}{\sum_{s}\lambda_{is}},
\end{equation}
where $\lambda_{is}$ is given by \eqref{intensity}. 

The true state $s = 1,2,\ldots, S$ at location $i=1,2,\ldots, R$ is then a realisation from a categorical distribution with probability $p_{is}$. 

This model specification for the ecological process used here is similar to the occupancy dynamics and encounter rate model used by \cite{spiers2022estimating} by assuming that occupancy probability equals one; and similar to the model used by \cite{wright2020modelling} by assuming Poisson counts with intensity $\lambda_{is}$ (refer to Table \ref{tab:generalisation} for the link between our model framework and that of \cite{spiers2022estimating} and \cite{wright2020modelling}).

\subsubsection{Classification process model}
For each observation, we assume there was imperfect classification and that the true state identities can be classified into $k = 1, 2, \ldots, K$ states. Assuming the true state distribution represents the true point pattern for the species under study, each reported information can be seen as a draw from any of the $K$ states under consideration with a given probability. As mentioned above, these states could be on any taxonomic level or include any unidentified group. For example, one could have four true states: common gull, herring gull, Audouin's gull and Sooty gull. These species can be reported in three states: large while-headed gulls, large black-headed gulls and other gulls. An example of the classification probability (confusion matrix) is shown in Table \ref{example}. 

\begin{table}[ht!]
\centering
\begin{tabular}{||c| c c c||} 
\hline
& & \textbf{Reported states} & \\

 \textbf{True states} & Large white-headed gulls & Large black-headed gulls & Other gulls \\ [0.5ex] 
 \hline\hline
Common gull & 0.8 & 0.1 & 0.1 \\ 
Herring gull & 0.9 & 0 & 0.1 \\
Audouin's gull & 0 & 0.9 & 0.1 \\
Sooty gull & 0 & 1 & 0 \\ [1ex] 
 \hline
\end{tabular}
\caption{Example of confusion matrix that applies to our model.}
\label{example}
\end{table}

 Let $\Omega_{sk}$ be the probability that a single true state $s \in \{1, \ldots, S \}$ is classified as state $k \in \{1, \ldots, K \}$. The probabilities across all the possible $k$ states sum to $1$. In studies with homogeneous classification probabilities, the confusion matrix for the classification can be expressed as:
\begin{equation}\label{omega_matrix}
\mathbf{\Omega} =  \begin{bmatrix}
\mathbf{\Omega}_{1}   \\
\mathbf{\Omega}_{2}  \\
\vdots \\
\mathbf{\Omega}_{S}  \\
\end{bmatrix}  =
 \begin{bmatrix}
\Omega_{11} & \Omega_{11}& \cdots & \Omega_{1k}& \cdots& \Omega_{1K}   \\
\Omega_{21} & \Omega_{22}& \cdots& \Omega_{2k}& \cdots& \Omega_{2K} \\
\vdots & \vdots & \cdots& \vdots & \cdots& \vdots\\
\Omega_{S1} & \Omega_{S2} & \cdots & \Omega_{Sk}& \cdots& \Omega_{SK} \\
\end{bmatrix}   
\end{equation}
where the rows correspond to the true states $s$ and the columns correspond to the reported states $k$. 

We model the heterogeneity in the classification probabilities by fitting a multinomial generalised linear model \citep{Fahrmeir2013} to each of the rows of $\mathbf{\Omega}$ defined in equation \eqref{omega_matrix}. We refer to this approach as the multiple multinomial generalised linear model (MMGLM, hereafter). We define the linear predictor of the MMGLM as:
\begin{equation}
\label{linear_predictor_classification}
\begin{split}
\zeta_{isk} &= \omega_{0sk}+ \sum_{j = 1}^{n_c} z_{ij}\omega_{jsk}\\
&= 
 \begin{bmatrix}
{\omega_0}_{11} & {\omega_0}_{12}&  \cdots& {\omega_0}_{1K}   \\
{\omega_0}_{21} & {\omega_0}_{22}&  \cdots& {\omega_0}_{2K} \\
\vdots & \vdots & \cdots& \vdots\\
{\omega_0}_{S1} & {\omega_0}_{S2} & \cdots& {\omega_0}_{SK} \\
\end{bmatrix}   + \ldots + \begin{bmatrix}
{\omega_n}_{11} & {\omega_n}_{12}&  \cdots& {\omega_n}_{1K}   \\
{\omega_n}_{21} & {\omega_n}_{22}& \cdots& {\omega_n}_{2K} \\
\vdots & \vdots &  \cdots& \vdots\\
{\omega_n}_{S1} & {\omega_n}_{S2} &  \cdots& {\omega_n}_{SK} \\
\end{bmatrix}* z_{in_c}  \\
\text{and in matrix notation:}\\
\boldsymbol{\zeta}_i & = \boldsymbol{\omega_0} + \sum_{j = 1}^{n_c}\boldsymbol{\omega_j} z_{ij} ;
\end{split}
\end{equation}
 where $\boldsymbol{\omega_{0}}$ is an $S \times K$ matrix of intercepts and $\boldsymbol{\omega_{j}}$ is an $S \times K$ matrix of covariate $j$ effects, and $z_{ij}$ is the $j^{th}$ covariate that drives the classification process at location $i$. Using equation \eqref{linear_predictor_classification} as the definition for the linear predictor, our estimates of the parameters $\omega_{0sk}$ and $\omega_{jsk}$ are identifiable with reference to one reported state. Therefore for each true state identity $s$, the classification probabilities for each reported state $k=1, 2, \ldots, K-1$ with reference to state $K$ is modelled as:
\begin{equation}\label{heterogenous}
    \ln \bigg(\frac{\Omega_{sk}}{\Omega_{sK}}\bigg) = ({\omega_0}_{sk} - {\omega_0}_{sK}) + ({\omega_1}_{sk} - {\omega_1}_{sK})*\textbf{z}_{1} + \ldots +({\omega_n}_{sk} - {\omega_n}_{sK})*\textbf{z}_{n};
\end{equation}
where $\textbf{z}_{1}$, $\ldots$, $\textbf{z}_{n}$ are covariate vectors.

This general framework has $S \times (K-1) \times (n_c +1)$ parameters to be estimated, where $S$ is the number of true states, $K$ is the number of reported states, and $n_c$ is the number of covariates in the classification process. Estimating these parameters can be very computationally expensive as the number of true states, reported states and covariates increases and would require significant numbers of misclassified state samples to estimate them. Therefore, we explored simplified forms of the generalised model in equation \eqref{heterogenous}.

A simplified case of equation \eqref{heterogenous} assumes that the classification covariate \textbf{z} only affects the probability of correctly classifying the states. In this instance, the matrix $\boldsymbol{\omega_j}$ has its off-principal diagonal elements $\omega_{jsk} = 0$ for $s \neq k$ for covariate $j$, and these parameters are not estimated (This is our study scenario "fixed-covariate" in Table \ref{tab:scenarios}). This simplification reduces the number of parameters estimated for the classification process by $n_c \times (K - S - 1)$, where $S$ is the number of true states, $K$ is the number of reported states and $n_c$ is the number of classification covariates. A further simplification would also assume that the average probability of correctly classifying the true state does not change with species, that is, $\omega_{0sk}$ is the same for all $s = k$ (This is our study scenario "fixed-intercov" in Table \ref{tab:scenarios}). The classification process heterogeneity is captured by the covariate effect $\omega_{1sk}$ for all $s=k$. The latter further reduces the number of parameters estimated by $S - 1$. This last simplification is useful especially when the states are very similar, and one would expect that their average classification probabilities would be the same.

Given which true state ($s$) was present at location $i$, the reported state at that location is a draw from $K$ states with probability $\mathbf{\Omega}$.
\begin{equation}
    Y_i|V_{i} = s \sim \text{Categorical}(\Omega_{is \cdot}),
\end{equation}
where $\Omega_{is.}$ refers to the probability of being classified as any of the $K$ states given that the true state identity at location $i$ is state $s$.

\subsubsection{Variable selection}
Fitting a complex model with these many parameters can result in an overfitted model. An overfitted model captures the pattern and noise in the training data, but performs poorly on validation or test data \citep{montesinos2022overfitting}. In some cases, the true state observation and classification process covariates may be highly correlated. These correlated covariates can inflate standard errors (reduce the precision) of the estimated parameters \citep{yu2015multicollinearity, roberts2017cross, caradima2019individual}. To avoid overfitting the model, there is a need to perform variable selection. In this study, we performed Bayesian variable selection, specifically the spike and slab prior to the classification process covariates \citep[for review of Bayesian variable selection see][]{o2009review}. For each of the classification process covariates, we re-define the linear predictor in equation \eqref{linear_predictor_classification} as:

\begin{equation} \label{variable selection}
\begin{split}
    \zeta_{isk} &= \omega_{0sk}+ \sum_{j = 1}^{n_c} \psi_{j}z_{ij}\omega_{jsk};\\
    \text{with} \quad \psi_{j} &\sim Bernoulli(q_j)
\end{split}
\end{equation}
where $\psi_{j}$ is the indicator that variable $j$ is selected with the expected probability $q_j$. The variable selection indicator specified in equation \eqref{variable selection} jointly selects the variable that affects all the true states in the model but can also be made state-specific \citep{ovaskainen_abrego_2020}. Probabilities $q_j$ closer to $1$ indicate that the variable contributes much to the model and should be selected, and those closer to $0$ indicate that the variable contributes less and can be discarded.  

\begin{table}[htbp!]
\tablestyle[sansbold]
\begin{tabular}{|p{0.16\linewidth}|p{0.15\linewidth}|p{0.3\linewidth}|p{0.39\linewidth}|}
\theadstart
  \thead Classification type &\thead Model & \thead True state observation process & \thead Classification process \\

\tbody
   Heterogeneous   & Variable / covariate & $\ln(\lambda_{is}) = \beta_{0s} + x_{1i}\beta_{1s}+ x_{2i}\beta_{2s}$ & $\ln(\frac{\zeta_{isk}}{\zeta_{isK}}) = \omega_{0sk} + \psi_{i} z_{1i}\omega_{1sk}$ \\ \hline
         & fixed-covariate & $\ln(\lambda_{is}) = \beta_{0s} + x_{1i}\beta_{1s}+ x_{2i}\beta_{2s}$ & $\ln(\frac{\zeta_{isk}}{\zeta_{isK}}) = \omega_{0sk} + \psi_{i} z_{1i}\omega_{1ss}$, \par $\text{where $\omega_{1sk} = 0 $ for $s \ne k$}$  \\ \hline
         & fixed-intercov & $\ln(\lambda_{is}) = \beta_{0s} + x_{1i}\beta_{1s}+ x_{2i}\beta_{2s}$ & $\ln(\frac{\zeta_{isk}}{\zeta_{isK}}) = \omega_{0sk} + \psi_{i} z_{1i}\omega_{1ss}$, \par $\text{where $\omega_{1sk} = 0$ for $s \ne k$}$ \par $\text{and $\omega_{0sk}$ is the same for $s=k$}$  \\ \hline
   Homogeneous      & intercept & $\ln(\lambda_{is}) = \beta_{0s} + x_{1i}\beta_{1s}+ x_{2i}\beta_{2s}$ & $\ln(\frac{\zeta_{isk}}{\zeta_{isK}}) = \omega_{0sk}$ \\ \hline
        &  constant & $\ln(\lambda_{is}) = \beta_{0s} + x_{1i}\beta_{1s}+ x_{2i}\beta_{2s}$ & $\Omega_{sk} \sim Dirichlet(\alpha_{sk})$,\par $\text{where $\alpha_{sk} \sim Exp(1)$}$ \\ \hline
         & main & $\ln(\lambda_{is}) = \beta_{0s} + x_{1i}\beta_{1s} + x_{2i}\beta_{2s} + \psi_{i} z_{1i}\beta_{3s}$ & $\Omega_{sk} \sim Dirichlet(\alpha_{sk})$, \par $\text{where $\alpha_{sk} \sim Exp(1)$}$  \\ \hline
 %
 %
 \tend
\end{tabular}
\caption{Variations in the MMGLM for the classification process defined by equation \eqref{variable selection} and observation process defined by equation \eqref{intensity} used as our study scenarios. There are three heterogeneous models: covariate, fixed-covariate and fixed-intercov, and three homogeneous models: intercept, constant, and main. }
\label{tab:scenarios}
\end{table}

\subsection{Modelling heterogeneity using ML prediction scores}
Some studies give weight to the true identities of the reported observations or state, for example, because they use machine learning to classify the observation. These prediction scores (such as $F_1$ score, mean square error, and logarithmic loss) are not  classification probabilities but are values that indicate how well the ML algorithm classifies data in the test sample. We can use this information to model the heterogeneity in the classification process and predict the true state identity as follows:

\begin{equation} \label{MLAlgorithm}
\begin{split}
 V_i| Y_i &\sim \text{Categorical} (p_{iks}); \\ 
 \text{with} \quad p_{iks} &= \frac{\lambda_{ik}w_{iks}}{\sum_{k} \lambda_{ik}w_{iks}}
\end{split}
\end{equation}
where $\lambda_{ik}$ is the average abundance of reported state $k$ at site $i$ and $w_{iks}$ is the predictive score of the $k^{th}$ reported state to true state $s$ at sampling unit $i$. 

It must be noted that this approach is a non-parametric approach to account for heterogeneity in the classification whereas the MMGLM is a parametric approach. Moreover, modelling the heterogeneity in the classification process using ML prediction scores models the covariate effects on the expected abundance of the reported states and corrects them using the prediction scores as weights to obtain the relative abundance of the true states. However, the MMGLM models the covariate effect on the expected abundance of true states and estimates the heterogeneity in the classification process using a parametric model. The prediction of the true state identity is done by weighing the expected intensity of true states with the estimated classification covariate.

\subsection{Generalisation of model framework}
The proposed framework generalises the existing mSDMs that account for misclassification in occupancy models. \cite{wright2020modelling} provided a framework to account for the homogeneity in the classification process, and our model is connected to this by using the relationship between the multinomial and the Poisson distribution \citep{steel1953relation} for the observation process as well as using a species by species constant model for the classification process (Table \ref{tab:generalisation}). \cite{wright2020modelling} further provided arguments that their proposed models were generalised forms of models for the binary detection of two species \citep{chambert2018two}, single species with count detections \citep{chambert2015modeling} and single species with binary detections \citep{chambert2018new}. Since our proposed framework can be seen as  a heterogeneous version of \cite{wright2020modelling}, our framework is also a generalisable form of the models in \cite{chambert2015modeling, chambert2018new, chambert2018two}. \cite{spiers2022estimating} provided an individual-level semi-supervised approach that estimates species misclassification with occupancy dynamics and encounter rates, and our model is connected to this if we assume a homogeneous classification process (that is $\boldsymbol{\omega_1} = \textbf{0}$), and assume that the occupancy probability is $1$ (Table \ref{tab:generalisation}).

\begin{table}[htbp!]
\tablestyle[sansbold]
\begin{tabular}{|p{0.2\linewidth}|p{0.4\linewidth}|p{0.4\linewidth}|}
\theadstart
\thead \centering Author &
\thead \centering Model Framework&
\thead \centering Link to our model \tabularnewline
\tbody
\cite{wright2020modelling} and the models their proposed framework generalises such as \cite{chambert2015modeling, chambert2018new, chambert2018two} &
\textit{True state observation process: absolute counts}

$z_{is} \sim \text{Bernoulli}(\psi_{is})$

$[V_{ijs}| z_{is} = 1] \sim \text{Poisson}(\lambda_{ijs})$
 &
\textit{True state observation process: relative abundance}

Choose $\psi_{is} = 1$, then $z_{is} = 1$ for all locations $i$ and true states $s$.

 $[V_{ijs}| z_{is} = 1] \sim \text{Poisson}(\lambda_{ijs})$ using the relationship between multinomial and Poisson distribution \citep{steel1953relation}.
 \\
&
\textit{Classification process:}

$[Y_{ijs}|V_{ijs} = v_{ijs}, z_{is} = 1 ] \sim \text{Multinomial}(v_{ijs}, \Omega_{sk})$, 

where $\Omega_{sk} \sim \text{Dirichlet}(\alpha)$ &

\textit{Classification process:}

$[Y_{ijs}|V_{ijs} = 1, z_{is} = 1 ] \sim \text{Categorical}(\Omega_{sk})$, 

where $\Omega_{sk}$ can be chosen as any of the homogeneous models described in Table \ref{tab:scenarios} \\ \hline

\multirow{ 2}{*}{\cite{spiers2022estimating}} &
\textit{True state observation process: occupancy dynamics and encounter rates}

     $z_{ist} \sim \text{Bernoulli}(\psi_{ist})$

$[V_{ijst}]  \sim Categorical \bigg(\frac{\lambda_{ijst}z_{ist}}{\sum_s \lambda_{ijst}z_{ist}} \bigg)$;
&
\textit{True state observation process: relative abundance}

 Choose $ t = 1, \psi_{ist} = 1$, then $z_{ist} = 1$ for all locations $i$ and true states $s$. 

$[V_{ijs}] \sim Categorical \bigg(\frac{\lambda_{ijs}}{\sum_s \lambda_{ijs}} \bigg)$;

\\
&
\textit{Classification process:}

$[Y_{ijs}|V_{ijs}] \sim \text{Categorical}(\Omega_{sk})$, 

where $\Omega_{sk} \sim \text{Dirichlet}(\alpha)$ &

\textit{Classification process:}

$[Y_{ijs}|V_{ijs} = 1] \sim \text{Categorical}(\Omega_{sk})$, 

where $\Omega_{sk}$ can be chosen as any of the homogeneous models described in Table \ref{tab:scenarios}.

\\
\tend
\end{tabular}
\caption{Extensions of our proposed models from homogeneous classification process studies done by \cite{wright2020modelling} and \cite{spiers2022estimating}. The table specifies the true state observation process model for \citeauthor{wright2020modelling} (absolute abundance model), \citeauthor{spiers2022estimating} (occupancy dynamics and encounter rate model) and ours (relative abundance model); and also the classification process model for \citeauthor{wright2020modelling} and \citeauthor{spiers2022estimating} (homogeneous classification process with classification probabilities simulated from Dirichlet distribution) and ours from heterogeneous models described in Table \ref{tab:scenarios}. Since our framework extends the work done by \citep{wright2020modelling}, it is safe to say that they are also generalised forms of \cite{chambert2015modeling, chambert2018new, chambert2018two}.  }
\label{tab:generalisation}
\end{table}

\subsection{Simulation Study} \label{simulation study}
\label{subsection:simulation study}
To demonstrate how our proposed model works and its use in prediction, we performed a simulation study for $S=2$ true states, and $K=3$ reported states over $R=1000$ sites. We simulated two covariates for the true state observation process and one for the classification, all from a Normal distribution with a mean of $0$ and variance of $1$. The true intensity was simulated from equation \eqref{intensity}. The intercepts of the model for the two true states were chosen as $\beta_{01} = -1$, $\beta_{02} = 0$ and covariate effect for each true state was chosen as $\beta_{11}= 4$, $\beta_{12}= -2$, $\beta_{21}= 0$ and $\beta_{22}= 0$ (that is, state $2$ is used as a reference category). The intercept and covariate effect for the classification process was chosen as follows:
\begin{equation*}\label{values}
\begin{split}
\omega_{0}&= 
 \begin{bmatrix}
2 & 0.5 & 0   \\
1 & 1 & 0 \\
\end{bmatrix}; \omega_{1} =\begin{bmatrix}
3 & -1 &   0   \\
-1 & 1 & 0 \\
\end{bmatrix}.
\end{split}
\end{equation*}

These values were chosen to obtain significant sample sizes of misclassified states. We simulated $200$ datasets with a heterogeneous classification process using the variable model in Table \ref{tab:scenarios} (referred to as the "\textbf{full model}"), $200$ with a homogeneous classification process by assuming $\boldsymbol{\omega_{1}} = \textbf{0}$ (a matrix of zeros) in equation \eqref{heterogenous} (referred to as "\textbf{reduced model}") and another $200$ with the covariate effect for the classification process (using the variable model in Table \ref{tab:scenarios}) that is correlated to the true state observation process covariate  (referred to as "\textbf{correlation model}"). The first simulated dataset explored modelling heterogeneity's effect on the classification process, whereas the latter explored the effect of having correlated covariates for the classification and true state observation process. The second assessed the effect of overfitting the classification process model (adding heterogeneity to the classification process when it should be homogeneous). Moreover, we assessed the impact of the number of misclassified samples on the mSDMs predictive performance. We increased the principal diagonal components of $\boldsymbol{\omega_{0}}$ by $6$ to obtain a reduction in the number of misclassified samples simulated.

We randomly withheld two hundred of the true state identities for each dataset simulated as our validation sample. The number of validation samples was not varied since \cite{spiers2022estimating} found that the number of validation samples had a modest effect on the model's predictive ability. We fitted the model under the various scenarios described in Table \ref{tab:scenarios} to the data and evaluated the model's predictive performance on the validation sample. 

\subsection{Case study: Gulls dataset} \label{case study}
The proposed model was used to analyse gull dataset downloaded from GBIF \citep{gbif}. The database hosts over two billion occurrence observations with well over a million observers (website visited on 17th February 2023). We were interested in the iNaturalist records since they have community verifications \citep{matheson2014inaturalist}. The observers collected these occurrence records and uploaded their observations with images and/or sounds that allowed for verification.  The reported observations go through iNaturalist community verification and are accepted as research grade when two-thirds of the community agreed to the taxon identification \citep{ueda2020inaturalist}, at which point they are published on GBIF. We assumed that the community-accepted taxon name is the true state \textbf{V}. We checked the iNaturalist platform to track the identification process of the observations and use the first reported identification as the reported state \textbf{Y}.

We obtained observations for some species of gulls in Denmark, Finland and Norway from 2015 to 2022. Specifically, we selected great black-backed gulls (\textit{Larus marinus}), herring (\textit{Larus argentatus}), common gulls (\textit{Larus canus}) and lesser black-backed gulls (\textit{Larus fuscus}) because the iNaturalist website reported that these species are commonly misclassified as the other. Any other species reported apart from the above-mentioned species were labelled as "others". We used annual precipitation \citep[accessed from the raster package; ][]{hijmans2015package} as true state observation process covariates in the model since it has been noted in some literature to affect the distribution of sea birds such as gulls \citep{jongbloed2016flight, algimantas2010abundance}.

The data obtained were presence-only records. Exploratory analysis revealed that there were no multiple observations at the same location for our selected species. We, therefore, assumed that our locations were discrete and treated the data as a marked point process, where the species that was reported at a location is given a value of $1$ and $0$ for the other species in this study. If we had a species list, then we could have modelled it as a repeated marked point process at the same location and treated the sites as a random effect.

Citizen science data are known to be affected by several sources of bias. Some common biases are spatial bias \citep{johnston2022outstanding, tang2021modeling}, and misclassifications \citep{johnston2022outstanding, tulloch2013realising}. We only accounted for the misclassifications in this study, as we are interested in explaining the classification process and not making inferences about the abundance of the gulls. Citizen scientists have been reportedly known to correctly classify species as they gain experience reporting the species \citep{vohland2021science}. We, therefore, modelled the variation in the classification process by using the number of reports made by each observer as a covariate in the classification process. If an observer has more than ten observations, the extra number of observations was calibrated at $10$. We used this number of observations for an observer as a measure of experience \citep{johnston2018estimates, kelling2015can}, although there are other indices for measuring effort or experience of the citizen scientist \citep{santos2021understanding, vohland2021science}. 

We also used ML algorithm to obtain prediction scores (specifically $F_1$ scores) for the possible true state identity of our downloaded data. The ML algorithm was a Convolutional Neural Network \citep[a modified form in ][]{koch2022maximizing} trained with data from all citizen science observations of any species in Norway. Since the ML algorithm is trained with all birds data from GBIF in Norway, we trained all our six study scenarios models summarised in Table \ref{tab:scenarios} with all data for our selected gull species in Norway and all data reported before 2022 in Finland and Denmark and used all data reported in 2022 in Finland and Denmark as our validation sample. The summary of the classifications (true and false positives) in the training and validation sample is presented in Supplementary information 2 (S2-2, S2-3).

\subsection{Fitting and evaluating the model}
We ran all the analyses using the Bayesian framework using the Markov chain Monte Carlo approach in the NIMBLE package \citep{de2017programming} using the R software \citep{Rsoftware}. We chose the priors for all true state observation parameters from a normal distribution with a mean of $0$ and standard deviation of $10$, and we chose the priors for the classification process parameters from Normal distribution with a mean of $0$ and standard deviation of $1$. For the scenarios: constant and main model, we chose the priors of the confusion matrix ($\Omega$) from the Dirichlet distribution with parameter alpha ($\alpha$), which has a prior of an exponential distribution with mean $1$.

We ran 3 chains, each with 10000 iterations; the first 5000 iterations were chosen as the burn-in. We check the convergence of the models by visually inspecting the trace plots and ignoring models with a Gelman-Rubin statistic \citep{brooks1998general} value greater than $1.1$. We kept a fifth of the remaining samples in each of the chains.

We used accuracy (the proportion of predicted true states identities from all the predictions of the validation samples), precision (the proportion of mismatched true states in the validation sample that were correctly classified from the predictions) and recall (the proportion of correct true states in the validation sample retrieved from the predictions) as performance metrics. We used a Bayesian approach and got the posterior distributions for the parameters. Higher values of the validation metrics indicated the preferred model. We also checked how well the model estimated the true state observation process parameters ($\boldsymbol{\beta_0}$, $\boldsymbol{\beta_1}$ and $\boldsymbol{\beta_2}$) and the classification process parameters ($\boldsymbol{\omega_0}$ and $\boldsymbol{\omega_1}$) by estimating the bias (difference between the true value and the estimated value) and precision of the parameters. 

\section{Results}

\subsection{Simulation study}


\subsubsection{Predictive performance}

\begin{figure}[ht!]
    \centering
\includegraphics{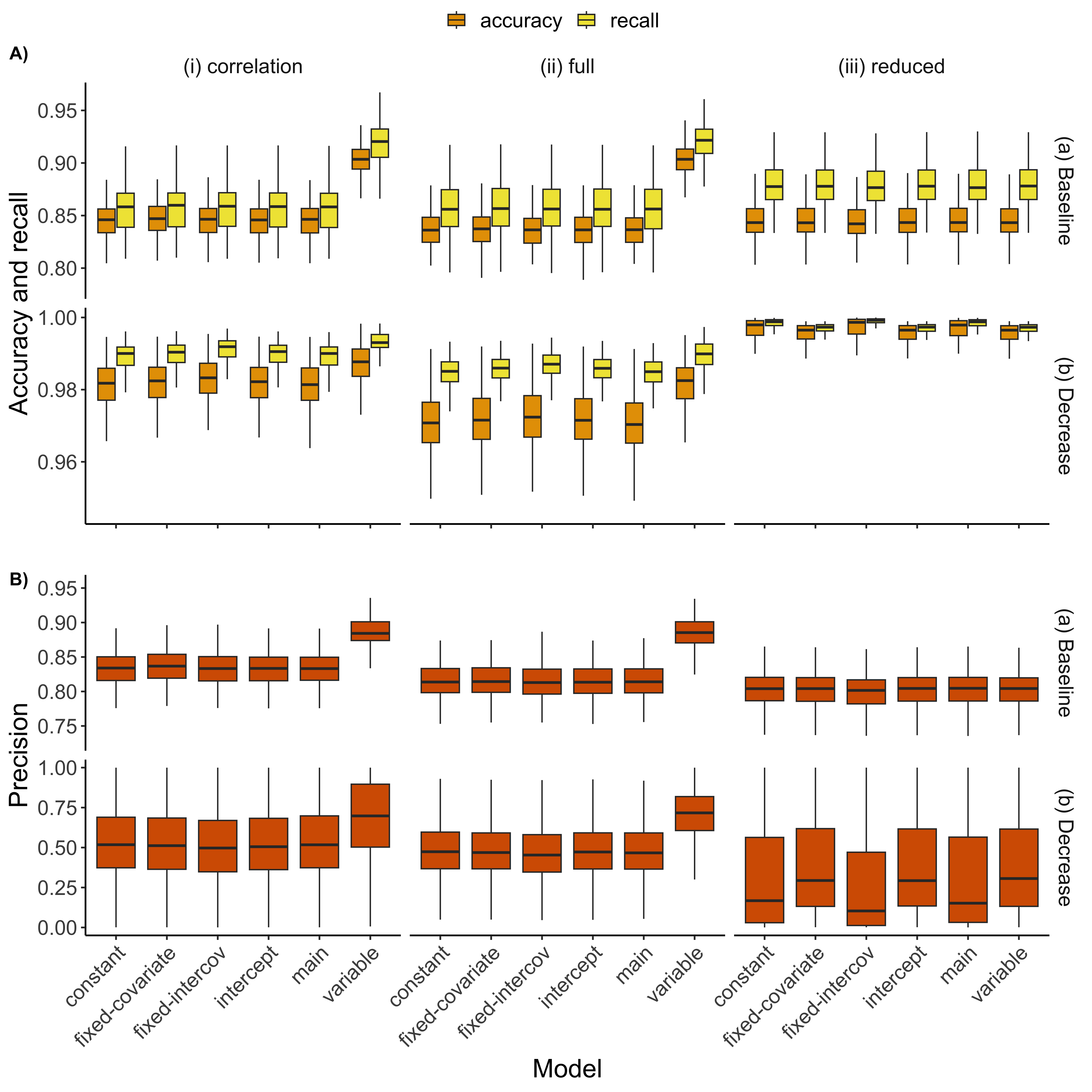}
\caption{Boxplot of validation metrics (accuracy, precision, recall) from the six study models defined in Table \ref{tab:scenarios} on the two hundred (200) withheld samples out of the thousand (1000) samples simulated in each dataset. Accuracy is the proportion of withheld samples that were correctly classified, recall is the proportion of correctly classified samples that were retrieved from the withheld samples, and precision is the proportion of the misclassified samples that were correctly classified. Each boxplot shows the median and the interquartile range (25 - 75\% quartiles). Each column shows the type of model used to simulate the dataset: "full" refers to using the variable/covariate model  in Table \ref{tab:scenarios}, "reduced" refers to using the intercept model in Table \ref{tab:scenarios} and "correlation" refers to using the variable model in Table \ref{tab:scenarios}, but with correlated true state observation and classification process covariates. The rows correspond to changes made to the number of misclassified samples in the simulated dataset: "Baseline" refers to using the values defined in section \ref{simulation study} and "Decrease" refers to reducing the number of misclassified samples by diagonal elements of $\mathbf{\omega}$ by 6 as described in section \ref{simulation study}}
    \label{fig:accuracy_and_precision}
\end{figure}

 We illustrated the gain in model performance by using the accuracy, recall and precision of our model's predictions (Figure \ref{fig:accuracy_and_precision} A and B). When data was simulated from the full and correlated models, there was a strong indication that the predictive performance of mSDMs improved when the variability in the classification process was included. That is the "variable" model performed best for the full and correlation models with the highest accuracy, recall and precision values (Figure \ref{fig:accuracy_and_precision} A (i - ii), B (i - ii)). The simplified heterogeneous models (fixed-intercov and fixed-covariate) however did not perform any better than the homogeneous models (Figure \ref{fig:accuracy_and_precision} A (i - ii), B (i - ii)). This suggested that simplifying the heterogeneous classification model did not yield any improvement in predictive performance, and the heterogeneous model that captures the entire variability (in this case variable model; Table \ref{tab:scenarios}) would be the best predictive model. When the classification process covariate was modelled as part of the observation process (main model), the model's predictive performance also performed similarly to the homogeneous models (Figure \ref{fig:accuracy_and_precision} A and B). 
 
 Overfitting a homogeneous classification process with a heterogenous one did not have any effect on the mSDM's predictive performance (Figure \ref{fig:accuracy_and_precision} A iii) and B iii)). We expected the overfitted heterogeneous models to have poor predictive performance \citep{montesinos2022overfitting}, but the heterogeneous and homogeneous performed similarly (with equal recall, accuracy and precision across all six study models). The Bayesian variable selection probability indicated that the homogeneous classification model was better (with the probability of including classification covariates in heterogeneous models being $0.359 \pm 0.012$; Supplementary information 2; S2-1). Although the simplified heterogeneous models did not yield improvement in predictive performance, they performed similarly to the variable model in the variable selection process.

\subsubsection{Effect of number of misclassified samples}
As we increased the number of misclassified samples in our simulated data, the precision increased by on average $30 \%$ and accuracy and recall reduced by $6 \%$ (Figure \ref{fig:accuracy_and_precision} A (i - ii), B (i - ii)). This decrease in accuracy and recall could be attributed to the reduction in the number of correct classifications in the simulated data as the number of misclassified samples increased. Moreover, the classification parameters were estimated better when the number of misclassified samples was higher, leading to the high precision of predictions (Supplementary information 2; S2-1). This suggested that our proposed model will be beneficial when one has a substantial number of misclassified samples.

\subsubsection{Bias in observation and classification process parameters}
 Although failure to account for misclassification in mSDMs can result in biased true state observation process parameters \citep{wright2020modelling, spiers2022estimating}, any method used to account for misclassification in mSDMs has a small effect on the accuracy and precision of the true state observation process parameters. The bias of the observation process parameters was consistently low for all six models, and the coverage was higher for all the scenarios under the full and reduced model (Supplementary information 2; S2-1). All the scenarios studied accounted for misclassification of some sort, thereby correcting for the bias in the observation parameters estimates \citep{wright2020modelling, spiers2022estimating}. The classification process parameters were estimated more accurately and precisely for the variable model than the other models (Supplementary information; S2-1). This was only possible in the case where we had enough misclassified samples. This suggests that if the objective of a study is to predict true state identity with mSDMs, then modelling the full heterogeneity can improve predictive performance; if the aim is inference on true state distribution, then heterogeneous models may not provide any advantage over homogeneous models. 

\begin{table}[htbp!]
\tablestyle[sansbold]
\begin{tabular}{*{5}{p{0.18\textwidth}}}
\theadstart
   \thead Method & \thead Accuracy & \thead Precision & \thead Recall & \thead Variable selection probability\\

\tbody
        Variable / Covariate & 0.97 & 0.1 & 0.99 & 0.71\\ \hline
        Constant & 0.97 & 0.1 & 0.99 & - \\ \hline
        Intercept & 0.97 & 0.1 & 0.99 & -\\ \hline
        Main & 0.97 & 0.1 & 0.99 & 0.29 \\ \hline
        Fixed intercov & 0.97 & 0.1 & 0.99 & 0.70\\ \hline
        Fixed covariate & 0.97 & 0.1 & 0.99 & 0.71 \\ \hline
        Machine Learning & 0.89 & 0.8 & 0.90 & - \\ \hline
 %
 %
 \tend
\end{tabular}
\caption{Validation metrics of the models under study on the withheld gull dataset. The accuracy is the proportion of correctly classified validated data, the precision is the proportion of mismatched identities that were correctly matched and recall is the proportion of correctly matched identities that were recovered. The number of validated samples was 384 out of which 10 were mismatched.}
\label{tab:data_validation}
\end{table} 

\subsection{Case study: Gull dataset}

All six study scenario models performed equally well regarding their predictive performance with high accuracy and recall but smaller precision (Table \ref{tab:data_validation}). The poor precision value could not be attributed to the insignificance of the classification covariate (observer experience) in explaining the heterogeneity in the classification process since the variable selection probabilities are closer to $1$ (Table \ref{tab:data_validation}) but to the small misclassification sample sizes (Supplementary Information 1; S2-2, S2-3).  However, the precision increased from 10\% to 80\% when the heterogeneity in the classification process was accounted for by using the prediction scores from the Machine learning algorithm (Table \ref{tab:data_validation}). The ML algorithm's prediction scores were individual observation-specific which provided direct information to the observation process model. However, the six classification models depended on the misclassified sample size to capture the heterogeneity in the classification process. This suggests that one remedy to improve mSDM's predictive performance for data with very small misclassified samples is to use ML weights to account for heterogeneity in mSDMs.

\begin{figure}[htbp!]
    \centering
\includegraphics{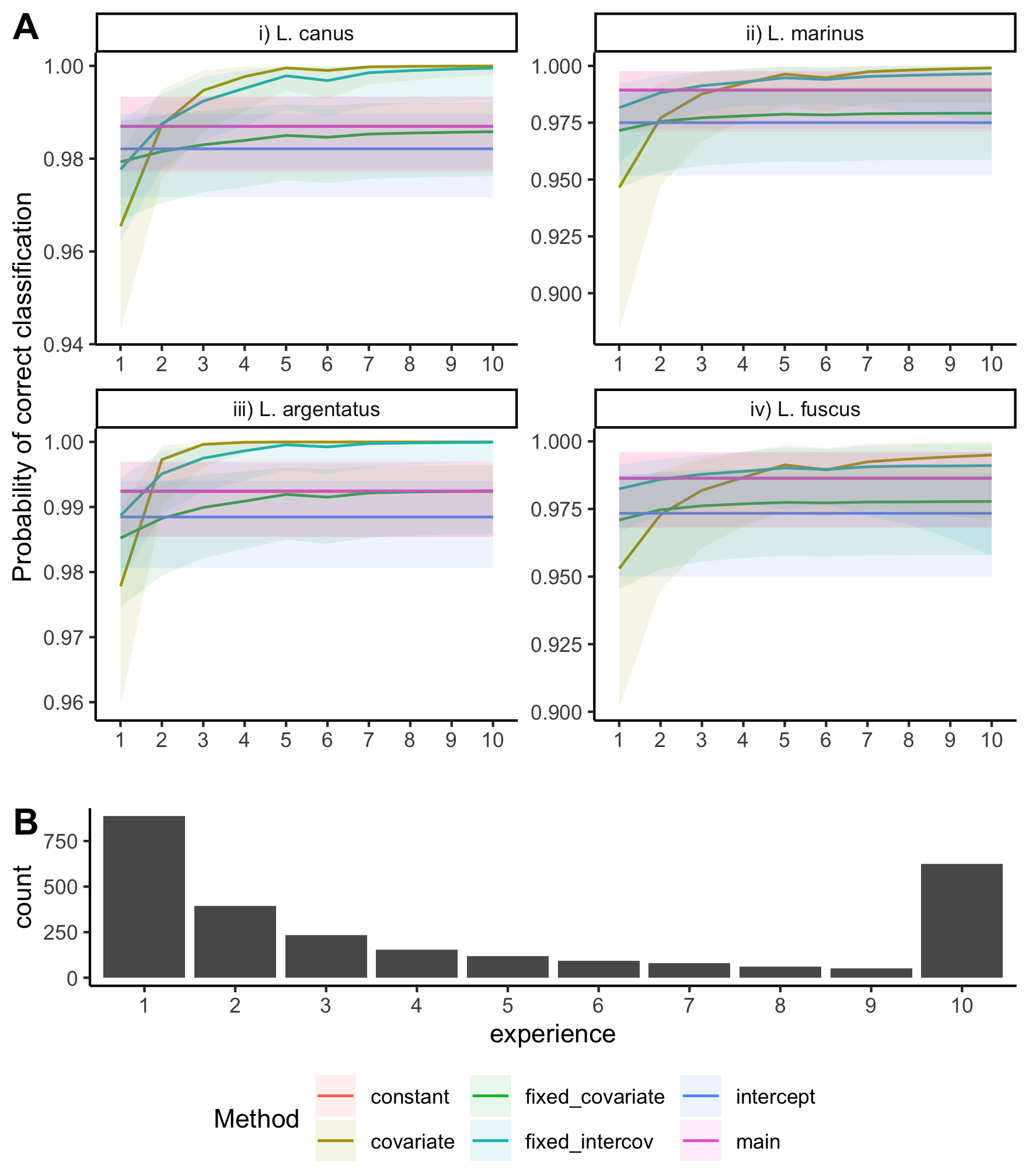}
\caption{Summary of results from the model fit to gull dataset showing A) Probability of correct classification for the common (\textit{Larus canus}), herring (\textit{Larus argentatus}), great black-backed (\textit{Larus marinus}) and lesser black-backed gulls (\textit{Larus fuscus}) and B) the distribution of the experiences of the observers used in the modelling. The ribbon around the correct classification probability estimates represents the 95 \% credible interval of the estimates.}
    \label{fig:omega}
\end{figure}

Although the study scenario models had smaller precision, it was observed that the probability of correctly classifying the gull species in Denmark, Finland and Norway increased with the experience of the observer (Figure \ref{fig:omega}). The pattern showed that observers have a high chance of making mistakes on their first few reports, and they get better as the number of reports increased \citep{vohland2021science}.

\section{Discussion}

The main objective of this paper was to propose a general framework to account for misclassifications from imperfect classifications (such as those from surveys) and uncertain classifications (from automated classifiers) in mSDMs. This work builds on previous work by \cite{spiers2022estimating, wright2020modelling} by accounting for the heterogeneity in the classification probabilities while allowing the classified categories to be more than the verified species (such as unknown species, morphospecies etc.). Moreover, we assessed the effect of overfitting a homogeneous classification process on the predictive performance of mSDMs and provided ways of checking the overfitting of the classification process model. 

Our study bridges the knowledge gap in the literature on accounting for misclassification in mSDMs by modelling the heterogeneity in the classification process. False positives are inevitable in biodiversity data \citep{miller2013determining, kery2020applied}. To model these false positives in this study, we presented the true state observation process as one model and the classification process as another model in a hierarchical form. We model the classification probabilities for each true state identity as a multinomial generalised linear model. This specification generalises the modelling of the classification process to model effects of covariates as fixed or random effects or both. For example, one can estimate the classification probabilities of each observer in volunteer-collected data by assuming a random observer effect. This formulation for the classification process also mitigates the modelling problems of using the Dirichlet distribution as the prior for the classification probabilities  \citep{spiers2022estimating}.

Furthermore, the specification of a separate state-space model for the true state observation process in the proposed framework allows the use of various multi-species observation models (such as joint species distribution models \citep{tobler2019joint, ovaskainen_abrego_2020}, Royle-Nichols model for abundance \citep{royle2003estimating}, amongst many others) to model the distribution of the true states. Our simulation study showed that the proposed model framework could estimate the true state relative abundance parameters (with the bias of estimated parameters close to zero; Supplementary information 2; S2-1), an observation noted in previous studies that use observation confirmation design to model misclassification in mSDMs \citep{spiers2022estimating, wright2020modelling, kery2020applied}. We have shown that the true state observation process model is a simplified form of occupancy and abundance models that account for misclassification (Table \ref{tab:generalisation}), so we believe our proposed framework can be extended to any design used to collect and verify data on the true states (for example, point processes, distance sampling, site-confirmation (multi-method) design, etc.), and any model used to fit the data (for example, multi-state occupancy model \citep{kery2020applied}, joint species distribution models \citep{tobler2019joint,ovaskainen_abrego_2020}).

Modelling the true state observation process with more complex models than the relative abundance models used in the study would add another level of hierarchical structure to the true state observation process (for example modelling detection probability or true occupancy state). This complexity could introduce confounding of the classification process and observation process parameters and with frequentist estimation approaches make the likelihood multi-modal \citep{kery2020applied}. This study did not explore such issues and further work can be done on this. A possible solution in the Bayesian framework to avoid such confounding issues would be to model the different processes with separate covariates, choose a good prior for the mSDM parameters, and use repeated survey visit data to model the true state observation process \citep{kery2020applied}. Moreover, the identifiability or confounding issues could also be tackled by using data with much information on detection and false positive detections such as those derived from acoustic surveys \citep{clement2022estimating} and integrating occupancy data that are not susceptible to misclassifications such as those from camera traps to those with misclassifications \citep{kery2020applied}.

Accounting for the heterogeneity in the classification process increases the predictive performance of mSDMs. The homogeneous classification models may sometimes be unable to explain the variation in the classification process \citep{conn2013accommodating}, leading to poor model predictive performance as a result of overfitting \citep{montesinos2022overfitting}. The simulation study showed a 30\% increase in precision and a 6\% decrease in accuracy and recall when the heterogeneity in the classification process was accounted for in the mSDMs (Figure \ref{fig:accuracy_and_precision} A, B). However, there was no change in predictive performance when a heterogeneous classification model overfitted a homogeneous classification process (Figure \ref{fig:accuracy_and_precision} A, B) due to the small classification covariate effect size, observed from the bias of parameter estimates and low Bayesian variable selection probability (Figure \ref{fig:accuracy_and_precision} A, B). Since the predicted posterior probability for the true state's identity heavily relies on the weights from the misclassification probability (Supplementary Information 1), failure to account for that would mean our posterior probability would be incorrectly estimated. The incorrectly predicted probability would lead to the underestimation of the prediction of the ranges of coverage and possibly abundance in the true states \citep{molinari2012monitoring}.

Fitting a more complex true state observation model with the covariate that explains the heterogeneity of the classification process does not provide enough information to improve the mSDM's predictive performance. Previous studies have shown that the estimates of the observation process model inform the estimation of the classification probabilities \citep{spiers2022estimating}, but the variability in the classification process cannot be inferred from variability in the observation process model (Figure \ref{fig:accuracy_and_precision} A (i - iii), Supplementary Information 1). Ecologists should therefore model the variability in the classification in its process model to gain the advantage in the mSDMs predictive performance.

Our model was parameterised with volunteer-collected gull data. These volunteer-collected data have several sources of bias in their generation, such as spatial bias, and misidentification of species, among many others. We acknowledge that all these sources of biases may be present in the data, but we only modelled the misidentification of species by using the number of previously collected data as a proxy measure for the observer's experience in the classification process model. The predictive performance of the homogeneous and heterogeneous models was approximately the same due to small misclassified samples (19 misclassified out of 1382 samples in training data (Supplementary Information S2-2) and 10 misclassified out of 378 samples in validation data (Supplementary Information S2-3)). However, the estimated covariate effect gives an idea of how the experience affects the probability of classifying a new observation. Specifically, the probability of correctly identifying the correct species increases with the observer's experience, as is noted in some literature \citep{santos2021understanding, vohland2021science, johnston2018estimates, kelling2015can}. Therefore, there is a trade-off between the model's ability to correctly classify mismatched data (precision) and understanding the covariate's effect driving the classification process when there are relatively small misclassified samples. 

The inclusion of ML prediction scores in the mSDMs to account for the heterogeneity in the classification process increased the precision of our predictions by 70\% (Table \ref{tab:data_validation}). These ML prediction scores are observation specific and provide much information about the classification process to increase the precision of the model. The information from the ML does not depend on the misclassified sample sizes but on the quality of the images \citep{koch2022maximizing}, making them advantageous to use in accounting for heterogeneity in the classification process when misclassified sample sizes are small (like we have in our gull data). 

The proposed model framework in this study is flexible and can be generalised into any species distribution model and integrated distribution model. The framework proposed fits into the frameworks provided by \cite{wright2020modelling} and \cite{spiers2022estimating} and any framework their study generalises. Our proposed classification process model, MMGLM, improved the predictive performance of mSDMs, but it heavily relies on the misclassified samples size. Furthermore, the confusion matrix defined in the model framework allows for the classification of different taxonomic groups, as opposed to just the species by species confusion matrix in \cite{wright2020modelling} and including morphospecies in the classification categories \citep{spiers2022estimating}. This will make it possible for citizen science data analysts to account for the misclassification of data at any level in the data collection process. We recommend that variable or model selection is performed during the analysis to check for overfitting. Moreover, ecologists should explore using ML prediction scores (where the prediction scores are available) as weights in mSDMs that aim at predicting true state distributions especially when the data has a small misclassified sample size.

\section{Data Availability}
The Gulls data used for this paper can be downloaded from GBIF (\url{https://doi.org/10.15468/dl.h24bp5}). All codes used for this paper are available at data dyrad with \url{https://doi.org/10.5061/dryad.0rxwdbs51}.

\section{Conflict of Interest}
The authors declare no conflict of interest. 

\section{Author Contribution}
All authors were involved in the idea conception and writing of the manuscript, KPA and RBO were involved with the methodology development, WK ran the ML algorithm and KPA ran the entire analysis used in this paper and led the writing of the manuscript.

\bibliographystyle{rss}
\bibliography{reference}
\end{document}